\documentclass{sig-alternate}

\begin{document}
%
% --- Author Metadata here ---
\conferenceinfo{EASE '15,}{April 27 - 29, 2015, Nanjing, China }
\CopyrightYear{2015} % Allows default copyright year (2015) to be over-ridden - IF NEED BE 
\crdata{978-1-4503-3350-4/15/04\dots\$15.00\\http://dx.doi.org/10.1145/2745802.2745817} 
% --- End of Author Metadata ---

\title {A Controlled Experiment for the Empirical Evaluation of Safety Analysis Techniques for Safety-Critical Software}

\numberofauthors{1} % in this sample file, there are a *total*

\author{
\alignauthor Asim Abdulkhaleq, Stefan Wagner \\
\affaddr{Institute of Software Technology, University of Stuttgart} \\
\affaddr{ Universit\"atsstra\ss e 38, 70569 Stuttgart, Germany} \\
\email{\{Asim.Abdulkhaleq, Stefan.Wagner\}@informatik.uni-stuttgart.de}
\\
}

\maketitle
% There's nothing stopping you putting the seventh, eighth, etc.
% author on the opening page (as the 'third row') but we ask,
% for aesthetic reasons that you place these 'additional authors'
% in the \additional authors block, viz.

% Just remember to make sure that the TOTAL number of authors
% is the number that will appear on the first page PLUS the
% number that will appear in the \additionalauthors section.

\maketitle
\begin{abstract}
\textbf{Context:} Today's safety critical systems are increasingly reliant on software. Software becomes responsible for most of the critical functions of systems. Many different safety analysis techniques have been developed to identify hazards of systems. FTA and FMEA are most commonly used by safety analysts. Recently, STPA has been proposed with the goal to better cope with complex systems including software. \textbf{Objective:} This research aimed at comparing quantitatively these three safety analysis techniques with regard to their effectiveness, applicability, understandability, ease of use and efficiency in identifying software safety requirements at the system level. \textbf {Method:} We conducted a controlled experiment with 21 master and bachelor students applying these three techniques to three safety-critical systems: train door control, anti-lock braking and traffic collision and avoidance. \textbf {Results:} The results showed that there is no statistically significant difference between these techniques in terms of applicability, understandability and ease of use, but a significant difference in terms of effectiveness and efficiency is obtained. \textbf {Conclusion:} We conclude that STPA seems to be an effective method to identify software safety requirements at the system level. In particular, STPA addresses more different software safety requirements than the traditional techniques FTA and FMEA, but STPA needs more time to carry out by safety analysts with little or no prior experience. 

\end{abstract}

\category{G.3}{Probability and Statistics/Experimental Design}{Controlled Experiments} 
\terms{Measurement, Experimentation}

\keywords{Controlled Experiment, Safety Analysis, STPA, FTA, FMEA}

\section{Introduction} 
At present, software is an integrated and complex part of many modern safety-critical systems and the amount of software is increasing. Thus, software safety must be considered to ensure system safety. Software can create hazards through erroneous control of the system or by misleading the system operators into taking inappropriate actions \cite{lev11}. Many accidents and losses have been caused by software, for example the loss of Ariane 5 \cite {LIO96}, Therac-25~\cite{LT93} and more recently the Toyota Prius. There exist over 100 safety analysis techniques which are used in industry \cite{ericson2005hazard}. Safety analysts, however, apply only few of them regularly. Recently, safety engineering literature contains more and more claims that the traditional safety analysis techniques are not adequate for analysing the current complex and fast-evolving systems \cite{lev2014, lev11}

Among traditional safety analysis techniques, FMEA (Failure Modes and Effects Analysis) and FTA (Fault Tree Analysis) are the most common techniques used in system safety and reliability engineering. They have been used at the component level but are less effective for factors that involve interactions between components, software flaws and external noise~\cite{lev11}. Indeed, major accidents in large complex systems are not caused by the failure of technical components, but rather software failures and organizational factors influence the design, manufacturing and operation of the system \cite{lev11}. A new trend is to advance the safety analysis techniques by system and control theory rather than reliability theory. STPA (Systems-Theoretic Processes Analysis) has been developed by Leveson since 2004 as a safety analysis approach based on STAMP (System-Theoretic Accident Model and Processes) for identifying system hazards and safety-related constraints necessary to ensure acceptable risk. 

\subsection{Problem Statement} There exists a plethora of safety analysis techniques which are currently used to identify the system safety requirements such as FTA, FMEA and more recently STPA. FMEA and FTA are widely used in industry for identifying hazards in safety-critical systems. STPA has been developed to identify hazards in today's complex systems and to overcome the limitations of applying the traditional safety analysis techniques on these complex systems. Since STPA is a relatively new technique, we have no clear understanding of its capabilities and difficulties compared to the existing safety analysis techniques in terms of extracting software safety requirements on the system level.\\

\subsection{Research Objectives} Our overall objective is to better understand different safety analysis techniques to aid safety engineers in practices. This study aims at exploring the application of three safety analysis techniques to get a full understanding about their capabilities in terms of identifying the software safety requirements of software-intensive systems at the system level.

\subsection{Contribution} The contribution presented in this paper is a quantitative comparison of the application of STPA vs.~FTA, STPA vs.~FMEA and FTA vs.~FMEA in gathering software safety requirements at the system level of three safety-critical systems: train door control, traffic collision avoidance and anti-lock braking. We made the comparison in a controlled experiment with 21 undergraduate and graduate students. 

\subsection{Context} The controlled experiment was conducted during the winter semester 2014/15 with master students from the international master programme (INFOTECH), who are studying in three different specializations: embedded systems engineering, computer hardware and software engineering and communication engineering, and bachelor students who are in their last year, studying in different departments: aerospace engineering, mechanical engineering and software engineering at the University of Stuttgart, Germany.

\section{Background}
In this section, we describe the three safety analysis techniques investigated in the experiment: FTA, FMEA and STPA. 

\subsection{Fault Tree Analysis}

FTA \cite{citeulike:7233882} was developed at Bell Laboratories in the early 1960's under a U.S. Air Force contract to analyse the Minuteman missile system. FTA is a top-down approach to identify critical failure combinations. It is based on the chain of event accidents model. It is widely used to discover design defects during the development of a system and to investigate the causes of accidents or problems that occur during system operations \cite{Robyn2005, Leveson:1982:SS:1005937.1005940, Leveson1983173}. The input of FTA is a known hazard, failure or accident, and a design description of the system under analysis. The FTA process can be divided into four main steps: 1) identify the root node (hazard or accident or failure); 2) identify the combination of events or conditions that caused the root node and combine them by using Boolean logic operators; 3) decompose the sub-nodes until events determined are basic (leaf nodes); and 4) identify minimum cut sets which are the smallest sets of basic events that cause the root node to occur.

\subsection{Failure Modes and Effects Analysis}

FMEA \cite{united1980procedures} was developed by NASA in 1971 and adopted by the U.S. military in 1974 as a systematic, proactive method for evaluating and discovering the potential failures, their potential cause mechanisms and the risks designed into a product or a process. FMEA helps to identify where and how the component might fail and to assess the relative impact of different failures. FMEA is, similar to FTA, based on the chain of events accidents model. FMEA is a bottom-up, structured, table-based process for discovering and documenting the ways in which a component can fail and the consequences of those failures. The input to FMEA is a design description of the system and component. The FMEA process can be divided into four sub-tasks: 1) establish the scope of the analysis, 2) identify the failure modes of each block; 3) determine the effect of each potential failure mode and its potential causes; and 4) evaluate each failure mode in terms of the worst potential consequences and assign the relative values for the assumed severity, occurrence and chance of detection to calculate the risk priority number. Ultimately, the analyst has to develop the recommended action required to reduce the risk associated with potential causes of a failure mode \cite{Robyn2005}. 
\subsection{Systems-Theoretic Process Analysis}
STPA (Systems-Theoretic Process Analysis) ~\cite {10.1109/TDSC.2004.1} is a safety analysis technique which is based on a Systems-Theoretic Accident Model and Process (STAMP) of accidents for large and complex systems. STPA has been developed to identify more thoroughly the causal factors in complex safety-critical systems including software design errors. With STPA, the system is viewed as interacting control loops and the accidents are considered as results from inadequate enforcement of safety constraints in the design, development and operation.

STPA aims to create a set of scenarios that can lead to an accident. It is similar to FTA but STPA includes a broader set of potential scenarios including those in which no failures occur but the problems arise due to unsafe and unintended interactions among the system components ~\cite{lev11}. STPA provides guidance and a systematic process to identify the potential for inadequate control of the system that could lead to a hazardous state which results from inadequate control or enforcement of the safety constraints. STPA is implemented in the three steps: 1) Establish the fundamentals of STPA before beginning the analysis by identifying system accidents or unacceptable loss events and draw the preliminary control structure diagram of the system; 2) identify the potentially unsafe control actions that could lead to a hazardous state; and 3) determine how each potentially hazardous control action could occur.
Recently, STPA has been applied in different areas in industry (e.g. Space Shuttle Operations \cite{Owens} and Interval Management in NextGen \cite{Fleming2013}). 
\section{Related Work}
There exist a number of empirical studies which have been performed to compare safety analysis techniques 

Jung et al.~\cite{6681355} conducted a controlled experiment and its replication to compare two safety analysis methods: Component Integrated Fault Trees (CFT) and Fault Tree (FT) with regard to the capabilities of the safety analysis methods (such as quality and the results) and to the participants' rating of the consistency, clarity and maintainability of these methods. The experiment was carried out with seven academic staff members working towards their PhD and then replicated with eleven domain experts from industry. The result showed that the CFT has potential of being beneficial for employees with little or no experience in fault tree analysis. CFT can be beneficial for companies looking for a safety analysis approach for a project using model-based development.

Mouaffo et al.~\cite{taibi_106} conducted two controlled experiments to compare fault-tree based safety analysis techniques: State Event Fault Tree analysis (SEFT) $vs.$ Dynamic Fault Tree (DFT) and SEFT $vs.$ Fault Tree combined with Markov Chains Analysis (MC). The two controlled experiments were conducted as a part of two lectures with students and researchers as participants. The first experiment (SEFT $vs. $ DFT) was run with eight students and six researchers (14 subjects). The second experiment (SEFT $vs.$ FT$+$MC) was conducted with twenty-seven students. The results showed that the subjects found DFT more applicable than SEFT and SEFT more applicable than FT+MC. Also, the subjects needed less time to perform DFT or FT+MC than to perform SEFT. 

Leveson et al.\ \cite{lev2014} compared STPA and the Aerospace Recommended Practice (ARP 4761) in a case study. The safety assessment process in ARP 4761 contains different safety analysis techniques such as Functional hazard analysis (FHA), Preliminary System Safety Analysis (PSSA), System Safety Analysis (SSA), FTA, FMEA and Common Cause Analysis (CCA). They compared the traditional safety analysis techniques which are recommended in ARP 4761 with STPA using the aircraft wheel brake system example. In particular, they compared FHA, PSSA, and SSA with STPA. The criteria of comparison between both methods were: underlying accident causality model, goals, results, role of humans in the analysis, role of software in the analysis. The results of their analysis of using both techniques on the same system showed that STPA identifies hazards omitted by the ARP 4761 process, particularly those associated with software, human factors and operations. 

To the best of our knowledge, there is no controlled experiment comparing the effectiveness, efficiency and applicability of FTA, FMEA and STPA in terms of identifying the software safety requirements at the system level. We have not found any experiment evaluating FTA $vs.$ FMEA or both $vs.$ STPA.

\section{Experimental Design}
In the following, we specify the goal of the experiment, describe the experimental design used for the experiment and the procedure followed for its execution.

\subsection{Study Goal}
In this study, we analyse the application of FTA, FMEA and STPA to safety-critical software to explore which technique is more effective, efficient, applicable, easier to use and understandable in the terms of identifying software safety requirements at the system level. 

\subsection {Experiment Variables} 

\subsubsection {Independent Variables} There are two independent variables in our experiment: safety analysis technique and safety-critical system. 
\subsubsection {Dependent Variables} Our experimental design consist of five dependent variables which are: applicability, understandability, ease of use, effectiveness and efficiency of the safety analysis techniques in identifying software safety requirements in the context of the whole system. \textbf{Applicability} is the measure of the degree to which the participants can apply the given safety analysis technique to software at the system level. \textbf{Understandability} measures the degree to which the participants understand the procedures and notations of the given technique. \textbf{Ease of use} measures the degree to which the participants find the application of the given technique to software easy. \textbf{Effectiveness} measures how many different software safety requirements identified by using one technique. \textbf{Efficiency} measures the amount of time needed by participants when they use each safety analysis technique to identify software safety requirements.

\subsection{Research Questions} 

To investigate the application of FTA, FMEA and STPA, the following research questions are addressed:\\
\textbf{RQ1: Which safety analysis technique is more applicable to identify software safety requirements at the system level?} RQ1 is relevant to investigate since the subject of FTA, FMEA, and STPA is the system, not software. Moreover, STPA was developed based on a system-theoretic accident model to overcome the lacking in ability of FTA and FMEA to address software and sub-systems interactions safety requirements. \\
\textbf{RQ2: Which notations and procedures of the safety analysis techniques are more understandable?} Easily understood notations and procedures of any safety analysis technique enable safety analysts to identify safety requirements easily. FTA, FMEA and STPA have specific procedures and notations that can be used to identify and document safety requirements, therefore, the understandability of their notations and procedures is important to investigate. \\ 
\textbf{RQ3: How difficult is it to use each safety analysis technique to identify software safety requirements?} This question is important to investigate since the successful adoption of any technique is dependent on how easy it is to use. The three techniques rely on different structures. The main structure of FTA is a fault tree, FMEA is a table and STPA is control structure diagram and tables. \\ 
\textbf{RQ4: How effective are the safety analysis techniques in identifying software safety requirements?} Each technique has strengths and limitations to address different safety requirements. In particular, traditional methods FTA and FMEA consider safety as component problem (e.g. component failure), whereas STPA considers safety as a system's control problem (e.g. hardware failure, software error, interactions between components failure) rather than a component failure problem. Therefore, we want to investigate their effectiveness in the terms of addressing software safety requirements. \\
\textbf{RQ5: How efficient are the safety analysis techniques in identifying software safety requirements?} Since FTA, FMEA and STPA provide different procedures and notations, therefore, different amount of time will be needed to perform safety analysis task by analysts. Therefore, we want to measure the amount of time needed for performing safety analysis tasks by using each technique. 

\subsection {Hypotheses}
The research questions RQ1$-$RQ3 were to investigate which safety analysis technique is more applicable, understandable and easy to use than others in identifying software safety requirements at the system level. The hypotheses $\textbf{H}_{1}$, $\textbf{H}_{2}$ and $\textbf{H}_{3}$ of the three dependent variables Applicability (A), Understandability (U) and Ease of use (E) were formulated respectively as follows: 
\textbf{Null hypothesis} $\textbf{H}_{n1,2,3j}$: There is no difference in the applicability, understandability and ease of use between safety analysis techniques to identify software safety requirements.\\
$ H_{n11}$: $\mu STPA_{A}$=$\mu FTA_{A}$, 
$ H_{n12}$: $\mu STPA_{A}$=$\mu FMEA_{A}$, \\
$ H_{n13}$: $\mu FMEA_{A}$=$\mu FTA_{A}$\\ 
$ H_{n21}$: $\mu STPA_{U}$=$\mu FTA_{U}$, 
$ H_{n22}$: $\mu STPA_{U}$=$\mu FMEA_{U}$, \\
$ H_{n23}$: $\mu FMEA_{U}$=$\mu FTA_{U}$\\ 
$ H_{n31}$: $\mu STPA_{E}$=$\mu FTA_{E}$, 
$ H_{n32}$: $\mu STPA_{E}$=$\mu FMEA_{E}$, \\
$ H_{n33}$: $\mu FMEA_{E}$=$\mu FTA_{E}$\\ 
\textbf{Alternative hypothesis} $\textbf{H}_{a1,2,3j}$: The participants will perceive the applicability, understandability and ease of use of safety analysis techniques differently.\\
$ H_{a11}$: $\mu STPA_{A}$>$\mu FTA_{A}$, 
$ H_{a12}$: $\mu STPA_{A}$>$\mu FMEA_{A}$, \\
$ H_{a13}$: $\mu FMEA_{A}$>$\mu FTA_{A}$\\
$ H_{a21}$: $\mu STPA_{U}$>$\mu FTA_{U}$, 
$ H_{a22}$: $\mu STPA_{U}$>$\mu FMEA_{U}$, \\
$ H_{a23}$: $\mu FMEA_{U}$>$\mu FTA_{U}$\\
$ H_{a31}$: $\mu STPA_{E}$>$\mu FTA_{E}$, 
$ H_{a32}$: $\mu STPA_{E}$>$\mu FMEA_{E}$, \\
$ H_{a33}$: $\mu FMEA_{E}$>$\mu FTA_{E}$\\
The research questions (RQ4) and (RQ5) were to investigate which technique is more effective and efficient in identifying the software safety requirements. The hypotheses $\textbf{H}_{4}$ and $\textbf{H}_{5}$ of the dependent variables: Effectiveness (F) and Efficiency (C) were formulated respectively as follows: \\
\textbf{Null hypothesis} $\textbf{H}_{n4,5j}$: There is no difference in the effectiveness and efficiency between the three safety analysis techniques. \\
$ H_{n41}$: $\mu STPA_{F}$=$\mu FTA_{F}$, 
$ H_{n42}$: $\mu STPA_{F}$=$\mu FMEA_{F}$, \\
$ H_{n43}$: $\mu FMEA_{F}$=$\mu FTA_{F}$\\
$ H_{n51}$: $\mu STPA_{C}$=$\mu FTA_{C}$, 
$ H_{n52}$: $\mu STPA_{C}$=$\mu FMEA_{C}$, \\
$ H_{n53}$: $\mu FMEA_{C}$=$\mu FTA_{C}$\\
\textbf{Alternative hypothesis} $\textbf{H}_{a4,5j}$: STPA is more effective and efficient than FMEA and FTA, and FMEA is more effective and efficient than FTA.\\
$ H_{a41}$: $\mu STPA_{F}$>$\mu FTA_{F}$, 
$ H_{a42}$: $\mu STPA_{F}$>$\mu FMEA_{F}$, \\
$ H_{a43}$: $\mu FMEA_{F}$>$\mu FTA_{F}$\\
$ H_{a51}$: $\mu STPA_{C}$>$\mu FTA_{C}$, 
$ H_{a52}$: $\mu STPA_{C}$>$\mu FMEA_{C}$, \\
$ H_{a53}$: $\mu FMEA_{C}$>$\mu FTA_{C}$ 
\subsection{Study Objects}
We used software controllers of three systems as study objects: Train Door Control System (TDCS), Anti-Lock Braking System (ABS) and Traffic Collision and Avoidance System (TCAS). 
\subsubsection {Train Door Control System (TDCS) }The train door control system \cite{Thomas2011, Sulaman:2014:PBR:2601248.2601260} was designed to open and close a door of train and monitors the status of the door. The main components of train door systems are: 1) door controller; 2) door actuator; 3) physical door; 4) and door sensor. Controller software in the door control unit continuously monitors all other components. The mechanism of the train door controller is that the sensor sends a signal about the door position and the status of the door to the door controller. Then, the door controller receives input from the door sensor with some other inputs from external sensors about the position of the train. The controller also gets an indication about possible emergencies from an external sensor. Then, the controller issues the door open or close commands. The actuator will receive these commands and apply mechanical force on the physical door.
\subsubsection {Anti-Lock Braking System (ABS)} ABS \cite{garrett2001motor} is an active safety feature designed to aid a car driver by preventing the wheels from completely locking during an emergency stop. ABS applies the optimum braking pressure to the individual wheels that ensure the vehicle can still be steered and shorten braking distances on slippery surfaces. ABS retains steering control and avoids skidding during an episode of heavy braking. It monitors the speed of each wheel to detect locking. 

The three main components of ABS are: 1) Wheel speed sensors that monitor wheel rotation speed; 2) hydraulic units that pump the brakes, and; 3) an electronic control unit (ECU) which receives information from the wheel speed sensors and, if necessary, directs the hydraulic units to pump the brakes on one or more of the wheels. 

\subsubsection {Traffic Collision and Avoidance System (TCAS)} TCAS \cite{James2004} is the system responsible for giving warning to pilots about the presence of another aircraft that may present a threat of collision. TCAS relies on a combination of surveillance sensors to collect data on the state of the intruder aircraft and a set of algorithms that determine the best maneuver that the pilot should make to avoid a mid-air collision. The main components of TCAS are: 1) TCAS computer unit processor; 2) antennas on aircraft body; and 3) TCAS presentation. The TCAS antenna continually surveys the airspace around an aircraft and transmits a signal. TCAS continuously calculates the tracked aircraft position and updates and displays the real-time position information on a display. 

\subsection {Participants}

We conducted the experiment with a total of 21 participants; 16 of them were master students from the international master's program in Information Technology (INFOTECH) with different specializations (e.g. embedded systems engineering, computer hardware and software engineering and communication engineering and media technology). Five of them were bachelor students from aerospace engineering, software engineering and mechanical engineering. We selected only the students who had a background and experience working with software in embedded systems. The participants were randomly assigned into three groups A, B and C. Each group has an equal number of participants (7 students).

\subsection{Experiment Instruments}

We give each group detailed guidelines which present the procedure of each technique with step-by-step instructions. We also give the participants the functional and design documentation for each study object before conducting the experiment. We provide the participants paper-based templates to document their results and time instead of using documentation applications (e.g. Word, Excel) or tool support for three techniques to avoid the influence of usability and ease or difficulty of the tools.

\subsection{Data Collection Procedure}
There are three different data sources that we use in this research: 
\subsubsection {Questionnaires} We developed questionnaires to help in gathering data from the participants. The questions were newly developed for testing the understanding, applicability, ease of use, and the effectiveness of each safety analysis technique to identify the software safety requirements for software-intensive systems at the system level. We divided the questionnaires into two categories: \emph{pre-experiment questionnaire} and \emph{post-experiment questionnaire}. We developed the pre-experiment questionnaire to capture information about the participants and their backgrounds. We developed the post-experiment questionnaire to gather the data after conducting each experiment. The post-experiment questionnaire includes 12 closed questions about how they worked with each safety analysis technique and what are their subjective assessments of each technique. The questionnaires sheets were distributed to participants during the experiment tasks.

\subsubsection{The final reports of safety analysis} The participants should report during their work the activities and the software safety requirements which are derived by using the given technique. We developed paper-based template to help the participants to document the software safety requirements at the end of their work. 

\subsubsection{Time sheets} Time sheets are a paper-based template for time tracking the participants use to record their amount of time spent during the case study work. The participants have to write down the starting and ending times for each task they performed.

\subsection{Measurements}
We use a 5-point Likert scale: "Strongly disagree", "Disagree", "Neither agree nor disagree", "Agree" and "Strongly Agree" as the response format of questionnaire items for rating most of questionnaire items which are relevant to the research questions RQ1, RQ2 and RQ3. For collecting data for research question RQ4, we use the final reports of the safety analysis techniques. The effectiveness of each technique can be computed with two measures: 1) the number of how many software safety requirements were found (recall); and 2) the number of how many different types of software safety requirements were found (coverage). To measure the coverage of each technique, we classify the software safety requirements into different types (e.g. missing input, wrong output or missing feedback) based on the classification of common software hazards discussed in the book \emph{NASA software safety} \cite{ NASA2004} and the common control flaws classification scheme introduced by Leveson in her book \emph{Engineering a Safer World} \cite{lev11}. To collect data for the question RQ5, we use the time tracking sheets to calculate the time needed by participants to perform each task in minutes.

\subsection{Experiment Design}
For designing the controlled experiment, we used a balanced factorial design with two factors of interest with three levels (three safety analysis techniques and three safety-critical systems). The factorial experiment consists of nine experimental units. Table 1 shows the balanced 3x3 two-factor factorial design matrix. We have a balanced design in which each group has the same number of participants. Each group has to work only one task adopting only one technique and one study object to reduce the influence of learning effects about the study objects or use the same technique in the experiment. Moreover, each group worked in the separated room to provide a good workplace for each group to avoid negative influence between the groups during the discussion. We chose to conduct the experiment as a blind experiment in which our participants have no idea about the hypotheses and the research questions to compare the results of the participants in each controlled experiment task.

The experiment was divided into three tasks and each task was conducted with all participants divided into three groups A, B and C (shown in Table 1): \textbf {Task 1}: The train door control system was the study object of this task. The level of complexity of this task was low; \textbf {Task 2}: The ABS system was used for the second task. The level of complexity of this task was medium; and \textbf {Task 3}: The traffic collision and avoidance system was the study object of this task. The level of complexity of this task was high.

% Table generated by Excel2LaTeX from sheet 'Tabelle1'
\begin{table}[htbdp]
\begin{center}
\caption{Experimental Design}
\renewcommand{\arraystretch}{1.2}
\scalebox{0.97}{\begin{tabular}{llccc}
& & \multicolumn{3}{c}{\textbf{Safety Analysis Techniques}} \\ \cline{3-5}
\multicolumn{1}{c}{\textbf{Tasks}} & \multicolumn{1}{c}{\textbf{Study Objects}} & FTA & FMEA & STPA \\
\hline
\multicolumn{1}{c}{Task 1} & \multicolumn{1}{c}{TDCS} & C & B & A \\

\multicolumn{1}{c}{Task 2} & \multicolumn{1}{c}{ABS} & B & A & C \\

\multicolumn{1}{c}{Task 3} & \multicolumn{1}{c}{TCAS} & A & C & B \\
\hline
\end{tabular}}%
\end{center}
\end{table}%

\subsection{Pilot Study}
After preparing the experimental design, we conducted a pilot study with 3 participants (1 student who is a developer of tool support for STPA and 2 Ph.D. students) from the research group Software Engineering, Institute of Software Technology, University of Stuttgart. The participants had a good background in software-critical systems and the STPA safety analysis technique. The pilot study allowed us to test our experimental design, to cover the potential problems, to confirm the selections of participants and the tasks, and to estimate the required time of the actual experiment. The pilot study was performed by following the same procedure of the experimental design. Since we had only 3 participants, we conducted the experiment with them as one group. After conducting the pilot study and gathering all necessary information, we tested the data analysis procedure to ensure that we can answer the research questions. In the following, we present learned lessons from analysing the pilot study.
The participants of the pilot study suggested the following improvements to our experiment design: 1) the presentation slides and tutorial guidelines should contain information on how to perform each task and what are the roles of each person; 2) we should organise a lecture on safety engineering for software-intensive systems before conducting the experiment to motivate the participants, increase their knowledge about software safety and introduce the terminology of safety engineering which are contained in the experiment; 3) the time for training and experiment for each group should be more than one hour and a half for each task; and 4) the instructions for the procedure for performing the safety analysis should be enhanced by including a concrete and complete example of safety analysis for each technique. 

Based on these suggestions from the participants in the pilot study, we improved our experimental design. The presentation slides were improved by explaining how the participants should work each task and what the role of each person. Each team has to select a moderator of the safety analysis team who will manage the team and select the person who is responsible for merging the results from each person, writing down the results of each step, and writing the common list of safety requirements at the end of the task (notary). The time of each training and experiment task work was fixed on average 4 hours per task ($3 \mbox {\ groups} \times 3 \mbox {\ techniques} \times 4 \mbox {\ hour} = 36 $ \mbox {\ hours}) for the experiment. 

\subsection{Study Preparation}

We invited each group to a lecture on safety engineering for software-intensive systems which lasted 1 hour ($3 \mbox {\ groups} \times 1 \mbox {\ hour} = 3$ hours). The lecture included an introduction about safety engineering, safety analysis techniques, what system safety and software safety are, how the software can contribute to accidents, what is different between software/system engineering and safety engineering and explanations about the safety engineering terminology such as hazard, accidents and undesired event (top event). At the beginning of the lecture with each group, we asked the participants to fill out the pre-experiment questionnaire to gather information about their background and pervious experience. At the end of the lecture, we asked the participants to choose the date and timeslot of training and experiment work. Each group had to select three timeslots during two weeks in the period 27th October 2014 to 10th November 2014. We had three exercise slots for the safety engineering lecture and nine exercise timeslots to conduct the experiment with the three groups.

Before conducting the experiment tasks work, we conducted an extensive preparation procedure with tutorials to ensure that our participants have a minimal level of knowledge regarding to the study objects and the application of the three safety analysis techniques. Each group received three training sessions on how to use these three safety analysis techniques: FTA, FMEA and STPA to identify the software safety requirements in the software-intensive systems. Furthermore, the participants were introduced to the theoretical part of each safety analysis technique and how to use the safety analysis technique to reduce the amount of risk of software to an acceptable level. We presented them during the training program a short video for each study object which explained the mechanism of the study object. 

The training procedure lasted 2 hours per group/technique ($3 \mbox {\ groups} \times 3 \mbox {\ techniques} \times 2 \mbox {\ hours} = 18 $ \mbox {\ hours}) for the three techniques. We divided the preparation time into two parts per technique: a half hour for introducing the theoretical part and one and a half hour for the practical exercises on using the safety analysis technique. 

\subsection{Execution Procedure}
After each training session, we ran each sub-task work of the three experiment tasks with a fixed time of 2 hours per sub-task work ($3 \mbox {\ groups} \times 3 \mbox {\ techniques} \times 2 \mbox {\ hours} = 18 $ \mbox {\ hours}) for the three tasks. At the beginning of each task, we explained the duties of the group, the study object, the safety analysis technique and the fixed time slot. We also provided the experiment task materials e.g. instructions of the given safety analysis technique, a detailed tutorial of the technique, which includes a detailed description of the study object and a practical example of safety analysis conducted using the given technique, the paper-based template and a time tracking sheet. Next, we asked the group participants to select a moderator and notary from them to manage the task session and record the final reports of the group. Then, we let the group work on the assigned sub-task. At the end of each sub-task, the team had to discuss and compare their individual software safety requirements to develop a common list of software safety requirements. Then the notary of the group had to fill in the paper-based template and all group participants had to answer a post-experiment questionnaire.

% Table generated by Excel2LaTeX from sheet 'Tabelle2'
% Table generated by Excel2LaTeX from sheet 'Tabelle2'
\begin{table*}[htbdp]
\begin{center}
\caption{Descriptive of Self-Assessed Participants' Background and Experience}
\renewcommand{\arraystretch}{1.14}

\begin{tabular}{rccccc}

& \multicolumn{5}{c}{\textbf{Descriptive Results}} \\ \cline{2-6}

\textbf{Self-Assessment on 5-point Likert scale} & Mode & Median & MAD & Min & Max 
\\ \hline

\multicolumn{1}{l}{Background in embedded systems software} & 3 & 3 & 0 & 1 & 4 \\
\multicolumn{1}{l}{Experience in embedded systems software} & 3 & 3 &1& 1 & 4 \\

\multicolumn{1}{l}{Background in safety analysis} & 2 & 3 &1& 2 & 4 \\

\multicolumn{1}{l}{Experience in safety analysis } & 2 & 2 & 1 & 2 & 3 \\
\hline
\end{tabular}%
\end{center}
\label{tab:addlabel}%
\end{table*}%
\begin{table*}[htbdp]
\begin{center}
\caption{Cronbach's Alpha of the Post-Experiment Questionnaire}
\renewcommand{\arraystretch}{1.14}
% Table generated by Excel2LaTeX from sheet 'Tabelle1'
\begin{tabular}{ lr r rrrrrrr}

\textbf{Dependent Variables} & \multicolumn{3}{c}{Applicatbility} & \multicolumn{3}{c}{Understandability} & \multicolumn{3}{c}{Ease of use} \\
\hline
\textbf{Technique} & FTA & FMEA & STPA & FTA & FMEA & STPA & FTA & FMEA & STPA \\ \cline{2-10}
\textbf{Cronbach's measure} & 0.72 & 0.72 & 0.79& 0.79 & \textbf{0.30} &0.71 & 0.74 & 0.77 & 0.87 \\
\hline
\end{tabular}%
\end{center}
\label{tab:addlabel}%
\end{table*}%

\subsection{Data Analysis Procedure}
For analysing the collected data and testing hypotheses, we used the following data analysis procedure:
\emph{Reliability Testing}: We calculate the Cronbach's Alpha to measure the reliability of the Likert-type scale

\emph {Descriptive Statistics}: We present the collected data with appropriate descriptive statistics to get a better understanding of the data. To answer RQ1--3, we first calculate the measures for central tendency of ordinal data of each question in post-experiment questionnaire mode, median and median absolute deviations (MAD) for its dispersion. Next, we compute the median for the related questions group of each dependent variable to perform descriptive statistics for each dependent variable. Then, we calculate the median of median, mode and MAD for the new values of the three variables. To answer RQ 4, we investigate the final report of each experiment and calculate the number of software safety requirements which are reported by the participants and we map them into different categories. Then, we calculate the recall and the coverage of each technique. They can be measured based on the following equations: 
\begin{equation}
\mbox{Recall (x)} = \frac{|\mbox{Number of SSRs}_{X }|}{| \mbox{Total of SSRs}_{ALL}| }
\end{equation}

\begin{equation}
\mbox{ Coverage (x)} = \frac{|\mbox{Number of SSRTs}_{X }|}{|\mbox{Total of SSRTs} _{ALL} |}
\end{equation}
Where \emph{x} acts the safety analysis approach, and \begin{math} SSRs_{X} \end{math} is the number of the software safety requirements that are reported during a safety analysis. \begin{math} SSRs_{ALL} \end{math} is the total number of safety requirements that are reported by all safety analysis approaches. \begin{math} SSRTs_{X} \end{math} is the number of different types of software safety requirements that are reported by using a safety analysis approach and \begin{math} SSRTs_{ALL} \end{math} is the total number of different software safety requirements categories that are addressed by all techniques. 

To answer RQ5, we calculate the time needed by participants to perform each task in minutes. 

\emph{Hypothesis Testing}: We use the Two-Way Analysis Of Variance (ANOVA) with Post-Hoc analysis for testing our hypotheses. We test the hypotheses at a confidence level of 0.05.

\section{Results and Discussion} 
We summarise the study population by analysing the per-experiment questionnaire answers. The participants represent a broad range of the system engineering domain. 38\% of participants study master of embedded systems engineering, 24\% computer hardware and software engineering, 14\% communication engineering and 14\% study bachelor of mechanical engineering, 5\% software engineering and 5\% aerospace engineering. Five of the respondents have a background in safety analysis, and they have prior knowledge on FTA (3 respondents) and FMEA (2 respondents) from their studies. Only one of the participants has a prior experience with both safety analysis techniques FTA and FMEA. Moreover, the participants rated their background and experience on an ordinal scales (1= very poor to 5= very good). The results are shown in Table 2. Overall, the participants reported a medium level of background and experience in the embedded systems software and low experience with safety analysis. The deviation in the participants' answers about their self-assessment in all responses was 1 or lower. Hence, we had a homogeneous group.

% Table generated by Excel2LaTeX from sheet 'Descriptive analysis'
\begin{table*}[htbp]
\begin{center}
\caption{Descriptive Statistics of the Post-Experiment Questionnaire}
\def\arraystretch{1.2}%
\renewcommand{\arraystretch}{1.1}
\begin{tabular}{rcrrrrrrrrrrrr}
\multicolumn{2}{c}{\textbf{}} & \multicolumn{3}{c}{\textbf{Applicability}} & \multicolumn{4}{c}{\textbf{Understandability}} & \multicolumn{5}{c}{\textbf{Ease of Use}} \\ \cline{3-14}
\multicolumn{1}{l}{\textbf{Technique}} & \textbf{Measure} & \multicolumn{1}{c}{\textbf{Q1}} & \multicolumn{1}{c}{\textbf{Q2}} & \multicolumn{1}{c}{\textbf{Q3}} & \multicolumn{1}{c}{\textbf{Q4}} & \multicolumn{1}{c}{\textbf{Q5}} & \multicolumn{1}{c}{\textbf{Q6}} & \multicolumn{1}{c}{\textbf{Q7}} & \multicolumn{1}{c}{\textbf{Q8}} & \multicolumn{1}{c}{\textbf{Q9}} & \multicolumn{1}{c}{\textbf{Q10}} & \multicolumn{1}{c}{\textbf{Q11}} & \multicolumn{1}{c}{\textbf{Q12}} \\
\hline
\multicolumn{1}{c}{FTA} & Median & 4 & 4 & 4 & 4 & 5 & 4 & 3 & 4 & 4 & 4 & 4 & 4 \\
\multicolumn{1}{c}{} & Mode & 4 & 4 & 4 & 4 & 5 & 5 & 2 & 3 & 4 & 4 & 5 & 4 \\
\multicolumn{1}{c}{} & MAD & 1 & 1 & 0 & 0 & 0 & 1 & 1 & 1 & 1 & 1 & 1 & 1 \\
\multicolumn{1}{c}{} & Minimum & 3 & 1 & 2 & 3 & 3 & 2 & 2 & 3 & 2 & 3 & 1 & 3 \\
\multicolumn{1}{c}{} & Maximum & 5 & 5 & 5 & 5 & 5 & 5 & 5 & 5 & 5 & 5 & 5 & 5 \\
\cline{2-14}\multicolumn{1}{c}{FMEA} & Median & 4 & 3 & 3 & 4 & 5 & 4 & 4 & 4 & 4 & 3 & 5 & 5\\
\multicolumn{1}{c}{} & Mode & 4 & 3 & 4 & 4 & 5 & 4 & 4 & 4 & 4 & 2 & 5 & 5 \\
\multicolumn{1}{c}{} & MAD & 0 & 1 & 1 & 1 & 0 & 1 & 1 & 0 & 1 & 1 & 0 & 0 \\
\multicolumn{1}{c}{} & Minimum & 1 & 2 & 2 & 2 & 4 & 2 & 1 & 1 & 2 & 2 & 2 & 2 \\
\multicolumn{1}{c}{} & Maximum & 5 & 5 & 5 & 5 & 5 & 5 & 5 & 5 & 5 & 5 & 5 & 5 \\
\cline{2-14}\multicolumn{1}{c}{STPA} & Median & 5 & 4 & 4 & 4 & 4 & 4 & 3 & 4 & 4 & 4 & 4 & 4 \\
\multicolumn{1}{c}{} & Mode & 5 & 4 & 4 & 4 & 3 & 4 & 3 & 4 & 4 & 4 & 3 & 4 \\
\multicolumn{1}{c}{} & MAD & 0 & 1 & 0 & 0 & 1 & 0 & 1 & 1 & 0 & 0 & 1 & 1 \\
\multicolumn{1}{c}{} & Minimum & 2 & 3 & 3 & 2 & 2 & 2 & 1 & 2 & 2 & 2 & 2 & 1 \\
\multicolumn{1}{c}{} & Maximum & 5 & 5 & 5 & 5 & 5 & 5 & 5 & 5 & 5 & 5 & 5 & 5 \\
\hline
\end{tabular}%

\end{center}

\label{tab:addlabel}%
\end{table*}%

\subsection{Descriptive Statistics\footnote{The questionnaires, the collected data and the analysis results are available on \protect \url{http://dx.doi.org/10.6084/m9.figshare.1275190}}}
We first calculated the reliability of the post-experiment questionnaire using the Cronbach's \begin{math} \alpha \end{math} reliability measure which is a common measure of internal consistency. Theoretically, a higher value of Cronbach's \begin{math} \alpha \end{math} is better (e.g. a value of 0.70 and above for the Cronbach's \begin{math} \alpha \end{math} is sufficient for internal consistency) \cite{Ritter2010}. Table 3 shows the reliability statistics measure for the dependent variables. All dependent variables for the three techniques demonstrated high reliability and internal consistency, ranging from 0.71 to 0.87 and they exceeded 0.70, except for understandability regarding FMEA. It was 0.30 which indicates a fairly good internal consistency.

Next, we calculated descriptive statistics for each question in the post-experiment questionnaire (shown in Table 4). In the following, we discuss in detail the results of the related questions of each dependent variable. 
\subsubsection {Applicability (RQ1)} Regarding the questionnaire, we asked the participants three questions Q1--Q3 about their opinions on the applicability of each technique. Q1 was on how the participants found the given safety analysis technique is applicable to identify relevant software safety requirements. The results of Q1 (Table 4) show that STPA received the highest number of \emph{Strongly Agree}, whereas FMEA and FTA received a moderate agreement. The deviation in the answers of this question was low with a MAD between 0 and 1. Q2 was about whether participants were able to identify the relevant software safety requirements by using the given technique. The results of Q2 denote that STPA and FTA have received a moderate agreement, whilst FMEA has received a neutral agreement. For all responses to Q2, the deviation was low (MAD: 1). Q3 was about whether the given technique provides a systematic way to identify software safety requirements. For Q3, the respondents agreed moderately with FTA, FMEA and STPA. The deviation was again between 1 and 0. 
\subsubsection {Understandability (RQ2)}
We asked the participants four questions Q4--Q7 to collect their opinions about the understandability of each technique. Q4 was on whether the participants found the procedure of a given technique easy to understand. For Q4 (Table 4), on average, we see a moderate agreement for the three procedures. The deviation here was low with a MAD between 0 and 1. Q5 was on how the graphical notations/tables of a given technique were easy to understand by participants during the experiment. From the results of Q5, we see that the notations/tables of FTA and FMEA have received a strong agreement, whereas the notations and tables of STPA received a neutral agreement. Q6 was on how the procedure of a given technique was clear and well-structured for participants. The results of Q6 show that respondents agreed strongly with the clarity of the procedure of FTA and moderately agreement with the procedure of FMEA and STPA. The deviation in all responses was 1 or lower. Q7 was on whether the participants were never confused when they used a given technique. Based on the results of Q7, FTA received moderate disagreement. The respondents mostly were neutral to STPA and moderately positive towards FMEA. The deviation here was also small (MAD: 1).

\subsubsection {Ease Of Use (RQ3)}
We asked participants five questions related to the ease of use: Q8--Q12. Q8 was on how easy to perform the major steps of a given technique were. The results of Q8 (Table 4) show that FMEA and STPA received a moderate agreement, whereas FTA received neutral responses. For all techniques, the deviation was small with a MAD between 0 and 1.\ Q9 was on how easy it was to identify software safety requirements by a given technique. By Q9, the respondents mostly agreed moderately with all techniques. The deviation here was also lower with a MAD between 0 and 1.\ Q10 was how easy to document the software safety requirements by the notations/tables of a given technique. The results of Q10 reveal that the respondents agreed moderately with FTA and STPA, whereas they disagreed moderately with FMEA. The deviation of all techniques was 1 or lower. Q11 was on how easy to draw the graphical notations/tables of a given technique. By Q11, the respondents agreed strongly with FTA and FMEA, whereas they were neutral with STPA. The deviation again was small with a MAD between 0 and 1. Finally, Q12 was on how easy it was to learn a given technique. The results of question Q12 show that the respondents agreed strongly with FMEA and moderately with FTA and STPA. The deviation in the answer was again low between 0 and 1. 

Based on the results in table 4, we computed the median of median of the related-questions values for the dependent variables: applicability (Q1--Q3), understandability (Q4--Q7) and ease of use (Q8--Q10) to achieve comparability between the dependent variables. Table 5 shows the median of median results of the dependent variables for each safety analysis technique.
We do not observe any significant differences between the three techniques regarding to applicability, understandability and ease of use. 

\subsubsection{Effectiveness and Efficiency (RQ4--5)} 
We calculated the recall and coverage of each technique based on the final report of each task. We also calculated the time needed for each technique to measure the efficiency. The effectiveness and efficiency were analysed for normality. Table 6 shows the descriptive statistics (mean, median, std. deviation) for the effectiveness and efficiency. The results show that the recall score has a mean of 0.443 for STPA, 0.326 for FMEA and 0.231 for FTA. The same order exists for the coverage score with a mean of 0.70 for STPA, 0.60 for FMEA and 0.30 for FTA. The results also reveal that the time needed (in minutes) has a mean of 88 for FTA, 94 for FMEA and 116 for STPA.

\subsection{Hypothesis Testing}
We performed Two-Way ANOVA with Post-hoc to check for statistically significant difference between the dependent variables of FTA, FMEA and STPA. Table 7 shows the ANOVA and Post-hoc tests results which we need for answering the set hypotheses H1--H5. 

For the hypotheses H1, the hypothesis test retains the null hypotheses with p-values ($p>0.05$): $ \textbf{H}_{n11} \mbox{\ and \ } \textbf{H}_{n13}$. Therefore, there is no statistically significant difference between the applicability of STPA $vs.$ FTA and FTA $vs.$ FMEA. However, the results reveal a statistical significant difference between the applicability of STPA and FMEA by rejecting the null hypothesis $ \textbf{H}_{n12}$ with the p-value 0.003. Thus, we accepted the alternative hypothesis $ \textbf{H}_{a12}$.

\begin{table}[htbp]
\begin{center}
\caption{The Median of Median Results for Applicability, Understandability and Ease of Use}
\def\arraystretch{1.1}%
\renewcommand{\arraystretch}{1.0}
% Table generated by Excel2LaTeX from sheet 'Median of Median data&Results'
\begin{tabular}{p{2.5cm}cr}
\multicolumn{2}{l}{\textbf{Dependent Variable ~Method}} & \textbf{Median of Median} \\ \hline
\multicolumn{1}{l}{\multirow{3}[6]{*}{Applicability}} & \multicolumn{1}{c}{FTA} & \multicolumn{1}{c}{4} \\
\multicolumn{1}{c}{} & \multicolumn{1}{c}{FMEA} & \multicolumn{1}{c}{3} \\
\multicolumn{1}{c}{} & \multicolumn{1}{c}{STPA} & \multicolumn{1}{c}{4} \\
\cline{2-3}\multicolumn{1}{l}{\multirow{3}[6]{*}{Understandability}} & \multicolumn{1}{c}{FTA} & \multicolumn{1}{c}{4} \\
\multicolumn{1}{c}{} & \multicolumn{1}{c}{FMEA} & \multicolumn{1}{c}{4} \\
\multicolumn{1}{c}{} & \multicolumn{1}{c}{STPA} & \multicolumn{1}{c}{4} \\
\cline{2-3}\multicolumn{1}{l}{\multirow{3}[6]{*}{Ease of Use}} & \multicolumn{1}{c}{FTA} & \multicolumn{1}{c}{4} \\
\multicolumn{1}{c}{} & \multicolumn{1}{c}{ FMEA} & \multicolumn{1}{c}{4} \\
\multicolumn{1}{c}{} & \multicolumn{1}{c}{STPA} & \multicolumn{1}{c}{4} \\
\hline
\end{tabular}%
\end{center}

\label{tab:addlabel}%
\end{table}%

% Table generated by Excel2LaTeX from sheet 'Time and Recall'
\begin{table}[htpb]
\begin{center}
\caption{Descriptive Statistics of Recall, Coverage and Time needed}
\def\arraystretch{1.2}%
\renewcommand{\arraystretch}{1.1}
\scalebox{0.97}{\begin{tabular}{rrcrc}
\multicolumn{2}{c}{\textbf{Factor~~Method}} & \textbf{Median} & \multicolumn{1}{c}{\textbf{Mean}} & \multicolumn{1}{c}{\textbf{Std. Deviation}} \\
\hline \multicolumn{1}{l}{\multirow{3}[6]{*}{Recall}} & \multicolumn{1}{l}{FTA} & 0.211 & 0.231 & 0.062 \\
\multicolumn{1}{l}{} & \multicolumn{1}{l}{FMEA} & 0.316 & 0.326 & 0.033 \\
\multicolumn{1}{l}{} & \multicolumn{1}{l}{STPA} & 0.455 & 0.443 & 0.038 \\
\cline{2-5}\multicolumn{1}{l}{\multirow{3}[6]{*}{Coverage}} & \multicolumn{1}{l}{FTA} & 0.300 & 0.400 & 0.173 \\
\multicolumn{1}{l}{} & \multicolumn{1}{l}{FMEA} & 0.600 & 0.633 & 0.153 \\
\multicolumn{1}{l}{} & \multicolumn{1}{l}{STPA} & 0.700 & 0.700 & 0.100 \\
\cline{2-5}\multicolumn{1}{l}{\multirow{3}[6]{*}{Time}} & \multicolumn{1}{l}{FTA} & 86 & 88 & 9.165 \\
\multicolumn{1}{l}{} & \multicolumn{1}{l}{FMEA} & 93 & 94 & 4.58 \\
\multicolumn{1}{l}{} & \multicolumn{1}{l}{STPA} & 115 & 116 & 4.04 \\
\hline
\end{tabular}}%
\end{center}
\label{tab:addlabel}%
\end{table}%

For the hypotheses H2 and H3, the hypothesis test reveals that there is no statistically significant difference between the ease of use and understandability of three techniques. Therefore, the null hypotheses $ \textbf{H}_{n21}$, $ \textbf{H}_{n22}$, $ \textbf{H}_{n23}$, $\textbf{H}_{n31}$, $ \textbf{H}_{n32}$, and $\textbf{H}_{n33}$ were retained with p-values ($p>0.05$).
For the hypotheses H4, the difference between the effectiveness of all three techniques was statistically significant. Therefore, the null hypotheses $ \textbf{H}_{n41}$, $\textbf{H}_{n42}$,$\textbf{H}_{n43}$ were rejected with p-values ($p<0.05$).

For the hypotheses H5, the test reveals that the difference between FTA and FMEA regarding efficiency was not statistically significant. Hence, the null hypothesis $ \textbf{H}_{n53}$ was retained with p-value ($ p>0.05$). Whereas the differences between STPA $vs.$ FTA, and STPA $vs.$ FMEA were statistically significant. Therefore, the null hypothesis \begin {math} \textbf{H}_{n51} \end{math} and \begin {math} \textbf{H}_{n52} \end{math} were rejected with p-values ($p<0.05$). 

\subsection{Discussion}
Based on the results of the experiment presented in this paper, we answered our research question as follows:
For RQ 1, the results show that the applicability of STPA to identify software safety requirements is better than the applicability of FMEA. On the other hand, we could not see a significant difference between the applicability of FTA $vs.$ STPA and FTA $vs.$ FMEA. For RQ2 and RQ3, we can see in table 7, there is no significant difference between FTA, FMEA and STPA in the terms of understandability and ease of use. That means the subjects perceived the understandability and ease of use of all three techniques as similar. 

To investigate RQ4, the results reveal that the subjects obtained a higher level of effectiveness when using STPA than FTA and FMEA. In particular, the average level of effectiveness obtained by the subjects that used STPA is 12\% higher than by the subjects that used FMEA and 21\% more than FTA. We also observed that the subjects achieved a better effectiveness level when using FMEA than FTA. In particular, FMEA outperformed FTA by 30\% in terms of coverage. In addition, the average level of effectiveness of FMEA is 9\% higher than FTA. 

To answer RQ 5, the results show that FTA outperformed FMEA and STPA in terms of efficiency. In fact, the subjects applying FMEA needed 6 minutes more on average than subjects applying FTA and the subject applying STPA needed 28 minutes more on average than subjects applying FTA. While the subjects applying STPA needed 22 minutes more on average than subjects applying FMEA. That means the application of STPA requires more time by subjects with little or no prior experience. The hypothesis test confirms the descriptive statistical results in the tables 5 and 6. Figure 1 shows Tufte\footnote{ \protect \url{http://www.edwardtufte.com/}}charts of recall, coverage and time needed. It can be seen that there is a difference in the recall, coverage and time needed between three techniques. \\

% Table generated by Excel2LaTeX from sheet 'Hypothesis'
\begin{table}[htbp]
\caption{Results of Hypothesis Test}
\def\arraystretch{1.0}%
\renewcommand{\arraystretch}{1.1}
% Table generated by Excel2LaTeX from sheet 'Hypothesis'
% Table generated by Excel2LaTeX from sheet 'Hypothesis'
\begin{center}

\scalebox{0.98}{ \begin{tabular}{p{2.5cm}rrr}
\multicolumn{1}{c}{\textbf{Variable}} & \textbf{Method (i)} & \textbf{Method (j)} & \multicolumn{1}{c}{\textbf{p-value }} \\
\hline
\multicolumn{1}{l}{\multirow{3}[6]{*}{Applicability }} & \multicolumn{1}{c}{\multirow{2}[4]{*}{FTA}} & \multicolumn{1}{c}{FMEA} & 0.090 \\
\multicolumn{1}{l}{} & \multicolumn{1}{c}{} & \multicolumn{1}{c}{STPA} & 0.185 \\
\cline{2-4}\multicolumn{1}{l}{} & \multicolumn{1}{c}{FMEA} & \multicolumn{1}{c}{STPA} & \textbf{0.003} \\
\hline
\multicolumn{1}{l}{\multirow{3}[6]{*}{Understandability}} & \multicolumn{1}{c}{\multirow{2}[4]{*}{FTA}} & \multicolumn{1}{c}{FMEA} & 0.675 \\
\multicolumn{1}{l}{} & \multicolumn{1}{c}{} & \multicolumn{1}{c}{STPA} & 0.347 \\
\cline{2-4}\multicolumn{1}{l}{} & \multicolumn{1}{c}{FMEA} & \multicolumn{1}{c}{STPA} & 0.176 \\
\hline
\multicolumn{1}{l}{\multirow{3}[6]{*}{Ease of Use }} & \multicolumn{1}{c}{\multirow{2}[4]{*}{FTA}} & \multicolumn{1}{c}{FMEA} & 0.707 \\
\multicolumn{1}{l}{} & \multicolumn{1}{c}{} & \multicolumn{1}{c}{STPA} & 0.137 \\
\cline{2-4}\multicolumn{1}{l}{} & \multicolumn{1}{c}{FMEA} & \multicolumn{1}{c}{STPA} & 0.064 \\
\hline
\multicolumn{1}{l}{\multirow{3}[6]{*}{Effectiveness}} & \multicolumn{1}{c}{\multirow{2}[4]{*}{FTA}} & \multicolumn{1}{c}{FMEA} & \textbf{0.044} \\
\multicolumn{1}{l}{} & \multicolumn{1}{c}{} & \multicolumn{1}{c}{STPA} & \textbf{0.001} \\
\cline{2-4}\multicolumn{1}{l}{} & \multicolumn{1}{c}{FMEA} & \multicolumn{1}{c}{STPA} & \textbf{0.021} \\
\hline
\multicolumn{1}{l}{\multirow{3}[6]{*}{Efficiency}} & \multicolumn{1}{c}{\multirow{2}[4]{*}{FTA}} & \multicolumn{1}{c}{FMEA} & 0.292 \\
\multicolumn{1}{l}{} & \multicolumn{1}{c}{} & \multicolumn{1}{c}{STPA} & \textbf{0.002} \\
\cline{2-4}\multicolumn{1}{l}{} & \multicolumn{1}{c}{FMEA} & \multicolumn{1}{c}{STPA} & \textbf{0.006} \\
\hline
\end{tabular} }%
\end{center}

\label{tab:addlabel}%
\end{table}%

\subsection{Threats To Validity}

This section presents the threats to validity of this study:
\emph{Construct validity} One possible construct validity is that the influence of the experimenter during the execution of the experiment. To reduce the problems which relate to this threat, we gave all relevant experiment materials and instruction to the participants before conducting the experiment. Furthermore, we tested our design with a pilot study to improve the experiment procedure, task description and the systems descriptions. 

\emph{Internal validity} The possible internal validity threat is that there were learning effects over time and the experiment tasks execution order. Regarding to these threats, we assigned participants randomly to the groups in which they can only apply one technique to one study object. Another internal threat is exchanging results and information between groups. We asked the groups to avoid discussion of their results with other groups. Moreover, our participants were not aware of the hypotheses of the experiment. They also were not evaluated on their performance. 

\emph{External validity} A possible external validity threat is that our participants were students and they may not be representative of professionals. We tried to have participants who already have knowledge about safety-critical systems. Most of our participants were master students who are not so far from being junior industry system engineers. A low number of participants can pose an external threat to generalise the results. 

\begin{figure*} 
\centering
\includegraphics[width=17cm , height=3.7cm ]{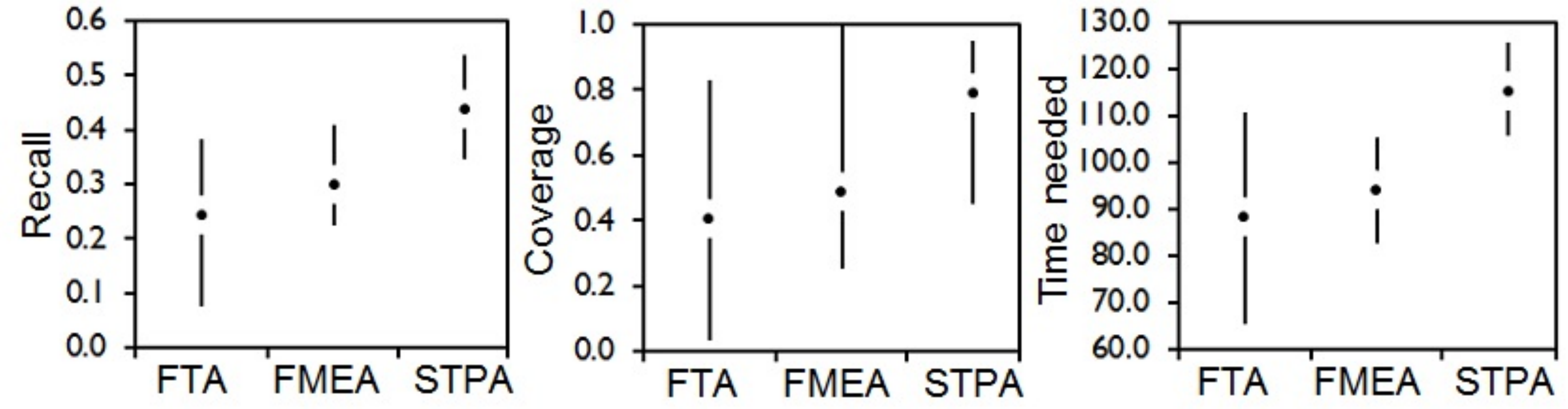}
\caption{Tufte Charts of Recall, Coverage and Time needed}
\label{fig:figure1}
\end{figure*}

\section{Conclusion \& Future Work} 
In this paper, we presented a controlled experiment aimed at comparing three different safety analysis techniques: FTA, FMEA and STPA with regard to their applicability, understandability, effectiveness and efficiency in terms of identifying software safety requirements of safety-critical software. We carried out the experiment with 21 graduate and undergraduate students at the University of Stuttgart. The subjects were trained in the safety analysis techniques during tutorial sessions. 

The results show that the subjects saw no difference between the applicability of FTA $vs.$ FMEA and FTA $vs.$ STPA, whereas they found STPA is more applicable then FMEA. In addition, the subjects experienced no significant difference between the understandability and ease of use of the three techniques. The results also show that STPA is more effective than FTA and FMEA, whereas, FMEA is more effective than FTA. Moreover, our results indicate that the participants applying FTA and FMEA needed nearly the same time. The participants applying FTA or FMEA needed less time than the participants applying STPA. That means FTA and FMEA seem to be more efficient than STPA.

We did not find any significant difference in the applicability, understandability and ease of use of the three techniques. However, STPA seems to be the most thorough method but it also needs the most time. Yet, for safety-critical systems, a high recall and coverage is probably more important. 

As future work, we plan to conduct a qualitative and quantitative empirical evaluation with safety analysis experts in industrial environments to improve the statistical significance and external validity of the experiment. 

\section{Acknowledgements}
The authors are grateful to Ivan Bogicevic, Jasmin Ramadani and Lukas Balzer for participating in the pilot study and for their many helpful comments and suggestions.

%\section{References}
%\bibliographystyle{acm}
\bibliographystyle{abbrv}

\bibliography{Literatur}
% That's all folks!

%\appendix

\end{document}